\documentclass[a4paper,fleqn,usenatbib]{mnras}
\usepackage[T1]{fontenc}
\usepackage{ae,aecompl}
\usepackage{graphicx}	% Including figure files
\usepackage{amsmath}	% Advanced maths commands
\usepackage{amssymb}	% Extra maths symbols
\usepackage{epstopdf}
\usepackage{xspace} 
\usepackage{natbib}
\newcommand{\spitzer}{\textit{Spitzer}\xspace}
\newcommand{\msun}{\hbox{$\hbox{\rm M}_{\odot}$}}
\newcommand{\herschel}{\textit{Herschel}\xspace}

\newcommand{\pc}{{\rm pc}}

\newcommand{\mum}{\micron\xspace}
\newcommand{\degree}{\mbox{$^{\circ}$}\xspace}

\title[]{Slingshot Mechanism for Clusters: Gas Density Regulates Star
  Density in the Orion Nebula Cluster (M42)}

\author[A. M. Stutz]{
Amelia M. Stutz$^{1,2}$\thanks{E-mail: astutz@astro-udec.cl, stutz@mpia.de}
\\
$^{1}$Departmento de Astronom\'{i}a, Universidad de Concepci\'{o}n,
Casilla 160-C, Concepci\'{o}n, Chile\\
$^{2}$Max-Planck-Institute for Astronomy, K\"onigstuhl 17, 69117 Heidelberg, Germany
}

\date{Accepted XXX. Received YYY; in original form ZZZ}

\pubyear{2017}

\begin{document}
\label{firstpage}
\pagerange{\pageref{firstpage}--\pageref{lastpage}}
\maketitle

\begin{abstract}
  We characterize the stellar and gas volume density, potential, and
  gravitational field profiles in the central $\sim$\,0.5~pc of the
  Orion Nebula Cluster (ONC), the nearest embedded star cluster (or
  rather, proto-cluster) hosting massive star formation available for
  detailed observational scrutiny.  We find that the stellar volume
  density is well characterized by a Plummer profile
  $\rho_{stars}(r) = 5755\,\msun\,\pc^{-3}\,(1+(r/a)^2)^{-5/2}$, where
  $a=0.36$~pc.  The gas density follows a cylindrical power law
  $\rho_{gas}(R) = 25.9\,\msun/\pc^3\,(R/\pc)^{-1.775}$. The stellar
  density profile dominates over the gas density profile inside
  $r\,\sim\,1$~pc.  The gravitational field is gas-dominated at all
  radii, but the contribution to the total field by the stars is
  nearly equal to that of the gas at $r\,\sim\,a$.  This fact alone
  demonstrates that the proto-cluster cannot be considered a gas-free
  system or a virialized system dominated by its own gravity.  The
  stellar proto-cluster core is dynamically young, with an age of
  $\sim$\,2-3~Myr, a 1D velocity dispersion of
  $\sigma_{\rm obs} = 2.6$~km~s$^{-1}$, and a crossing time of
  $\sim$\,0.55~Myr.  This timescale is almost identical to the gas
  filament oscillation timescale estimated recently by Stutz \& Gould
  (2016).  This provides strong evidence that the proto-cluster
  structure is regulated by the gas filament.  The proto-cluster
  structure may be set by tidal forces due to the oscillating
  filamentary gas potential.  Such forces could naturally suppress low
  density stellar structures on scales $\gtrsim\,a$.  The analysis
  presented here leads to a new suggestion that clusters form by an
  analog of the "slingshot mechanism" previously proposed for stars.
\end{abstract}

\begin{keywords}
open clusters and associations: individual: M42 (ONC) - 
Stars: formation - 
Infrared: stars -
ISM: clouds - 
Clouds:  Individual: Orion A 
\end{keywords}

\section{Introduction}

The Orion Nebula Cluster (ONC, also known as M42) is the nearest site
of massive star formation and significant embedded cluster available
for detailed observational scrutiny.  At a distance of $\sim$\,400~pc
\citep[e.g.,][]{menten07,sandstrom07,schlafly14,kounkel17} it has been
the subject of many studies \citep[e.g.,
][]{jones88,hillenbrand98,kroupa99,kroupa00,kroupa01,hartman07,tobin09,megeath12,megeath16,meingast16,dario16,port2016}
scrutinizing in detail the stellar and protostellar content of the
region.  Generally, these studies have restricted themselves to the
investigation of the stellar content of the cluster for lack of high
quality observations that reliably trace the total gas mass
distribution, which have only been obtained recently
\citep[][]{stutz15,stutz16}.

\begin{figure*}
  \includegraphics[trim = 8mm 22mm 0mm 30mm,clip]{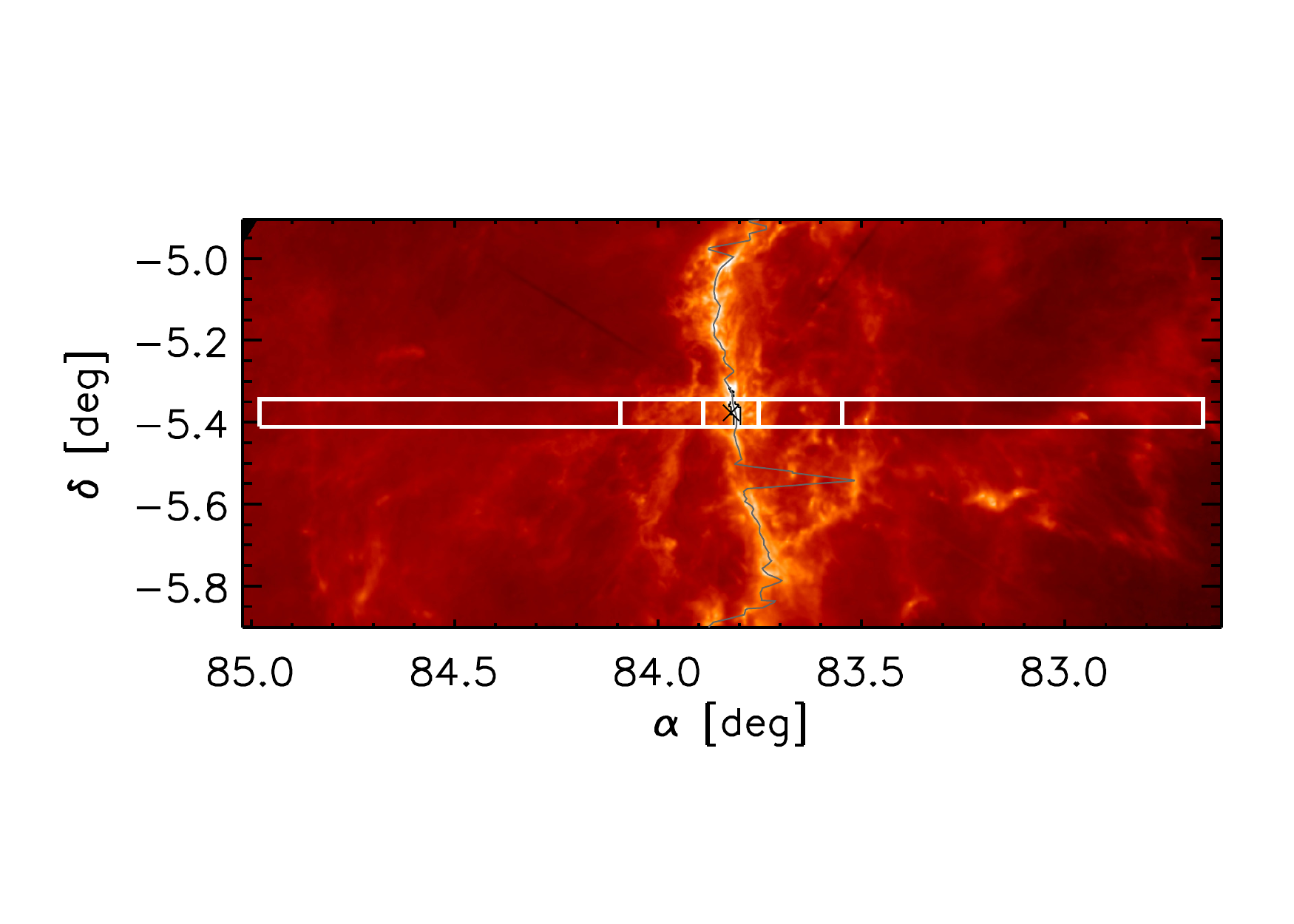}
  \caption{N(H) map from \citet{stutz15} of the Integral Shaped
    Filament (ISF, including M42).  The vertical scale of the figure
    is 7.3~pc, the horizontal scale is $\sim$\,17~pc.  The saturated
    map region within which we apply the APEX 870\micron\ correction
    is indicated by the black contour (see text for details). The grey
    line shows the dust ridgeline \citep{stutz16}. The $\times$-symbol
    indicates the stellar center of mass of the ONC. The large white box
    is 0.5~pc $\times$ 17~pc in size while the inner vertical lines
    indicate radii of 0.5~pc and 2.0~pc.}
    \label{fig:isf_cor}
\end{figure*}
%%star center: 
%%racen1 =        83.820818
%%deccen1 =       -5.3772089
%%racen1 = 05:35:17.00
%%deccen1 = -05:22:37.95
%%RA offset between dust ridgeline and cluster center [arcsec] = -39.1

The ONC is directly associated with a larger high line-mass gas
filament called the Integral Shaped Filament \citep[ISF;
][]{bally87,tatematsu08,stutz16,kainulainen17}.  The ONC is partially
embedded in the high density filament and lies slightly in foreground
\citep[e.g., ][]{odell01,wen95}, as is apparent from the continuous
distribution of stellar extinctions (S.\ T.\ Megeath, private
communication 2017).  The filament, with a wave-like morphology, has
dimensions of $\sim$\,7.3~pc in height and a horizontal oscillation
semi-amplitude of $\sim$\,1.5~pc \citep{stutz16}.  The stellar
distribution is highly elongated as the stars follow the gas
\citep[e.g.,][]{hillenbrand98,dario14,megeath16,stutz15,stutz16}. Indeed,
significant effort has been dedicated to characterizing the
asymmetries in the projected stellar distribution \citep[e.g.,][and
references above]{megeath16}.  Nevertheless, the relationship between
the stellar and gas volume density structures in the central
$\sim$\,0.5~pc of the ONC has so far remained unexplored.

Recently, \citet{stutz16} proposed, based on a high density of
observations, that the gas filament in the ISF is oscillating and
ejecting young stars \citep[the ``Slingshot'', see
also][]{boekholt17,schleicher17}.  Given the extremely close
association of the ONC stars with the gas filament, this scenario
immediately raises the question of whether such filamentary gas
oscillations might affect the structure of the embedded stellar
cluster.  Here we investigate this possibility through analysis of
both the stellar \citep{megeath16} and gas \citep{stutz15,stutz16}
distributions in the center of the ONC.

\begin{figure*}
  \includegraphics[trim = 0mm 0mm 0mm 0mm, clip]{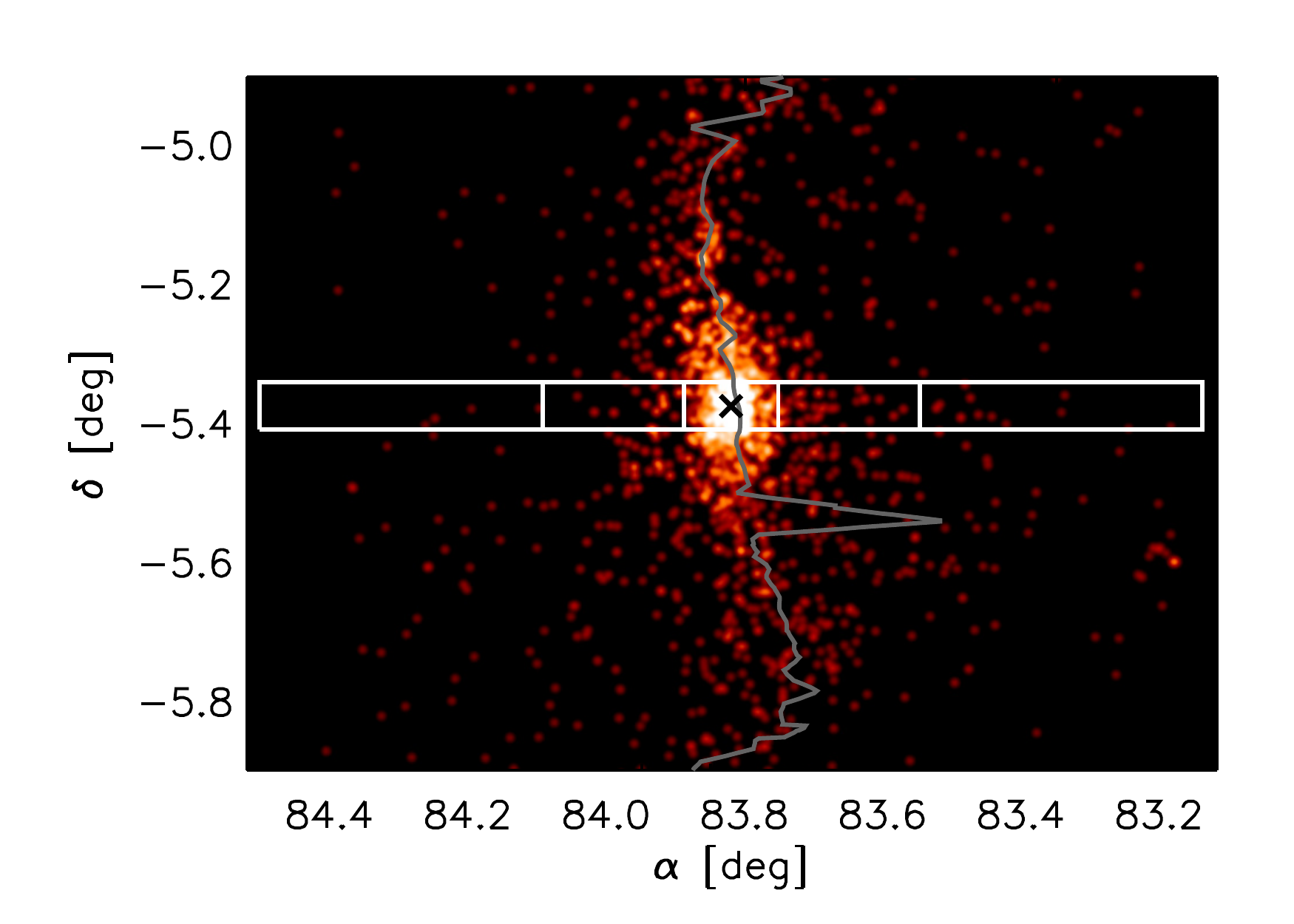}
  \caption{Star mass map of the ISF.  The vertical scale of the figure
    is 7.3~pc, the horizontal scale is $\sim$\,10~pc.  The
    $\times$-symbol indicates the stellar center of mass of the ONC at
    $\alpha=$~05:35:17.0, $\delta=$~$-$05:22:37.95. The dust ridgeline
    is indicated as a dark-grey line.  The white box is
    0.5~pc\,$\times$\,10~pc in size. The large white box is 0.5~pc
    $\times$ 10~pc in size while the inner vertical lines indicate
    radii of 0.5~pc and 2.0~pc. }
    \label{fig:isf_stars}
\end{figure*}
%% contour levels: 1500 & 750 msun/pc^2 

Previous efforts at simulating open cluster formation have focused
mainly on proto-cluster formation within turbulent molecular clouds
\citep[e.g.,][]{bate09,fujii2016}.  Here, proto-clusters (or embedded
clusters) are distinct entities from clusters.  Both are defined by
concentrations of stars whose gravity is sufficiently strong to
influence the stellar dynamics.  However, in clusters, the gravity is
dominated by the stars whereas in proto-clusters it is dominated by
the gas out of which the stars are still forming.  With this analysis
we provide the essential ingredients necessary for simulating cluster
formation conditions in ONC-analogs, in which clusters form on massive
dynamical gas filaments.  We quantify both the stellar and gas volume
density distributions using simple geometric assumptions and discuss
the relationship between them. We find that the gas and star
gravitational field profiles reach near-equality at
$r = a = 0.36$\,pc, the softening scale of the stellar density
profile.  At all other radii the gas dominates the gravitational
field.  This, combined with the fact that the cluster crossing time is
very similar to the estimated timescale for the gas filament motions,
strongly suggests that the cluster profile and dynamics are controlled
by the gas filament.

\section{Gas and stellar mass maps}

\subsection{Gas mass map}

We use the column density map from \citet[][; see also Stutz \& Gould
2016]{stutz15}.  This map was derived from the \herschel dust emission
data at 160~\mum, 250~\mum, 350~\mum, and 500~\mum.  The final
resolution of the map is about 20\arcsec.  We refer the reader to
\citet{stutz15} for further details.

The \citet{stutz15} N(H) map was corrupted due to saturation in the
center of the Orion Nebula Cluster (ONC) over a small elongated region
covering less than 1.5\arcmin$\times$6\arcmin.  We correct this defect
using the APEX 870~\mum data \citep[e.g.,][]{stanke10,stutz13}.  This
correction assumes that: (1) the spatial filtering in the ground-based
submillimeter data is negligible across the narrow extent of the
\herschel N(H) map artifact (with a maximum width of
$\sim1.5$\arcmin); and (2) the 870~\mum emission is optically thin and
traces the same dust emission as the \herschel maps.  We use the
region immediately outside the saturation artifact to scale the
870~\mum data in E-W strips to fill in the missing column density
information.

Figure~\ref{fig:isf_cor} shows the corrected N(H) map of the Integral
Shaped Filament (ISF) region, with M42 and the forming star cluster
(ONC) located in the center.  The grey line in
Figure~\ref{fig:isf_cor} is our recalculated dust ridgeline \citep[for
details see][]{stutz16} based on the corrected N(H) maps, which
follows the maximum N(H) as a function of $\delta$.  It is interesting
to note the spike in the dust ridgeline near $\delta = -5.5\degree$,
caused by discontinuity in the filament that causes the ridgeline to
"jump" to the to the nearest N(H) maximum, in this case to the west of
the main filament.  The filament also appears discontinuous in the
high density gas tracer N$_2$H$^+$ \citep{tatematsu08,hacar17},
indicating that while the filament appears approximately like a single
structure, it is in fact being broken apart at this location
immediately below the cluster formation site (see discussion).

\subsection{Stellar mass map}

We use the disk star count map of Orion~A based on the 
\spitzer data analyzed in detail by \citet{megeath16} \citep[see
  also][]{megeath12}.  Figure~\ref{fig:isf_stars} shows the stellar
mass map.  Briefly, in our map each young star is represented by a
delta function and then convolved to a 37\arcsec\ beam size
(approximately matching the \herschel 500~\mum beam). The weight of
each delta function is equal to the inverse estimated local incompleteness at
the source location.  In regions of high nebulosity such as the center
of the ONC the completeness correction fails due to the limited amount
of information in the images. Therefore \citet{megeath16} use the COUP
x-ray data \citep{feigelson05} to augment the incomplete \spitzer
counts.  To accomplish this, \citet{megeath16} apply weighting factors
to account for two principal effects:
\begin{itemize}
\item[1.] The COUP data detect a different source population than the
  \spitzer data: \spitzer is based on IR excess, while the COUP data
  are sensitive to all xray emitting stars.
\item[2.] The \spitzer data are incomplete due to nebulosity (see above).
\end{itemize}
In order to account for these effects and match the two data-sets,
\citet{megeath16} compare the nebulosity-corrected \spitzer counts to
the COUP surface densities in the region of overlap where both the
\spitzer and COUP incompleteness are minimized (that is, outside of
the core of the ONC) to scale the COUP profile to the \spitzer counts.
In the regions where the \spitzer data do not provide complete
information, the scaled COUP counts are used.  This analysis is
described in detail in \citet{megeath16}.  The total number of stars
with disks across Orion~A is 5265.  The total nuber of stars with
disks in the ISF (shown in Fig.~\ref{fig:isf_stars}) is 3344, while
the total number of stars within the box centered on the ONC (see
Fig.~\ref{fig:isf_stars}) is 1031.  The \spitzer and COUP counts can
in principle also be used without a nebulosity correction, and the
resulting mass distribution differences are discussed below.

%%***********
%%YSO counts in the region:
%%YSOs in all Orion A:       5265.5117
%%YSOs in ISF:       3343.8839
%%YSOs in orange slice (from separate code): 1031.4 
%%***********

To convert the young star count map into a total stellar mass map we
must assume two quantities: 1.) a mean stellar mass of
0.5~\msun\ \citep{kroupa01} , and 2.) a disk fraction of 0.75
\citep{megeath16}.  Figure~\ref{fig:isf_stars} shows the projected
stellar mass distribution of the ISF.  Ideally, we would base the
conversion from star counts to stellar mass on empirically observed
quantities, such as direct measurements of the IMF.  \citet{dario12}
measure the ONC IMF over a region covering about
30\arcmin$\times$30\arcmin.  Based on various evolutionary tracks
they measure the (model dependent) characteristic mass (the peak IMF
mass) to be about 0.3~\msun\ (see their Table~4).  As they note, this
relatively high characteristic value is driven by their finding that
the low mass population is deficient compared to other regions.  They
argue that incompleteness at low masses is not driving this
result. Such values would correspond to somewhat higher mean stellar
masses than those assumed here.  Nevertheless, given the status of the
current observational evidence, our assumed mean stellar mass is
justified.  Furthermore, our basic ONC results are not affected by
small shifts in the assumed mean stellar mass as these will not change
the shape of the fitted density profile (see below).  Our disk fraction
assumption of 0.75 is analyzed in detail in \citet{megeath16}.  We
refer the reader to that work (specifically Sect.~2.2) for discussion
on the derivation of the disk fraction.  We note that \citet{getman14}
argue for a higher disk fraction in the center of the ONC. However,
small shifts in this number do not affect our results (see below).
Here we follow \citet{megeath16} and assume a single average disk
fraction throughout our map. This approach works well in regions
dominated by disk stars, such as the ONC.

\begin{figure}
  \includegraphics[width=\columnwidth]{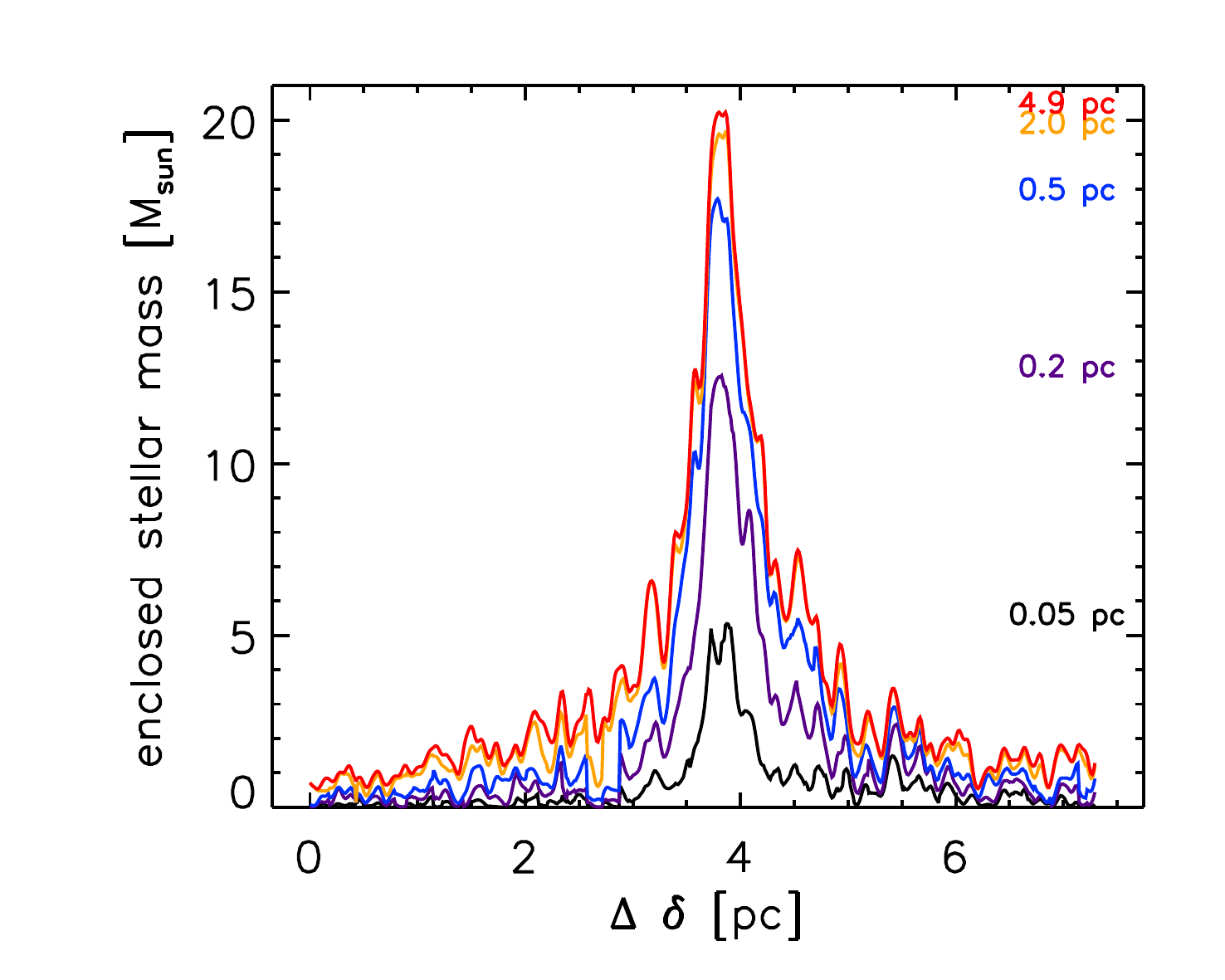}
  \caption{Enclosed stellar mass at a given declination starting from
    the Southern edge of ISF (at Dec. $= -5.9\degree$) to the northern
    edge (Dec.$ = -4.9\degree$).  Different curves indicate the mass
    enclosed within various projected radii from the dust ridgeline.
    The maximum extent of the stellar mass distribution is about 5~pc
    although most of the stellar mass lies within 2~pc of projected
    distance from the dust ridgeline.}
  \label{fig:smassd}
\end{figure}

\begin{figure}
  \includegraphics[width=\columnwidth]{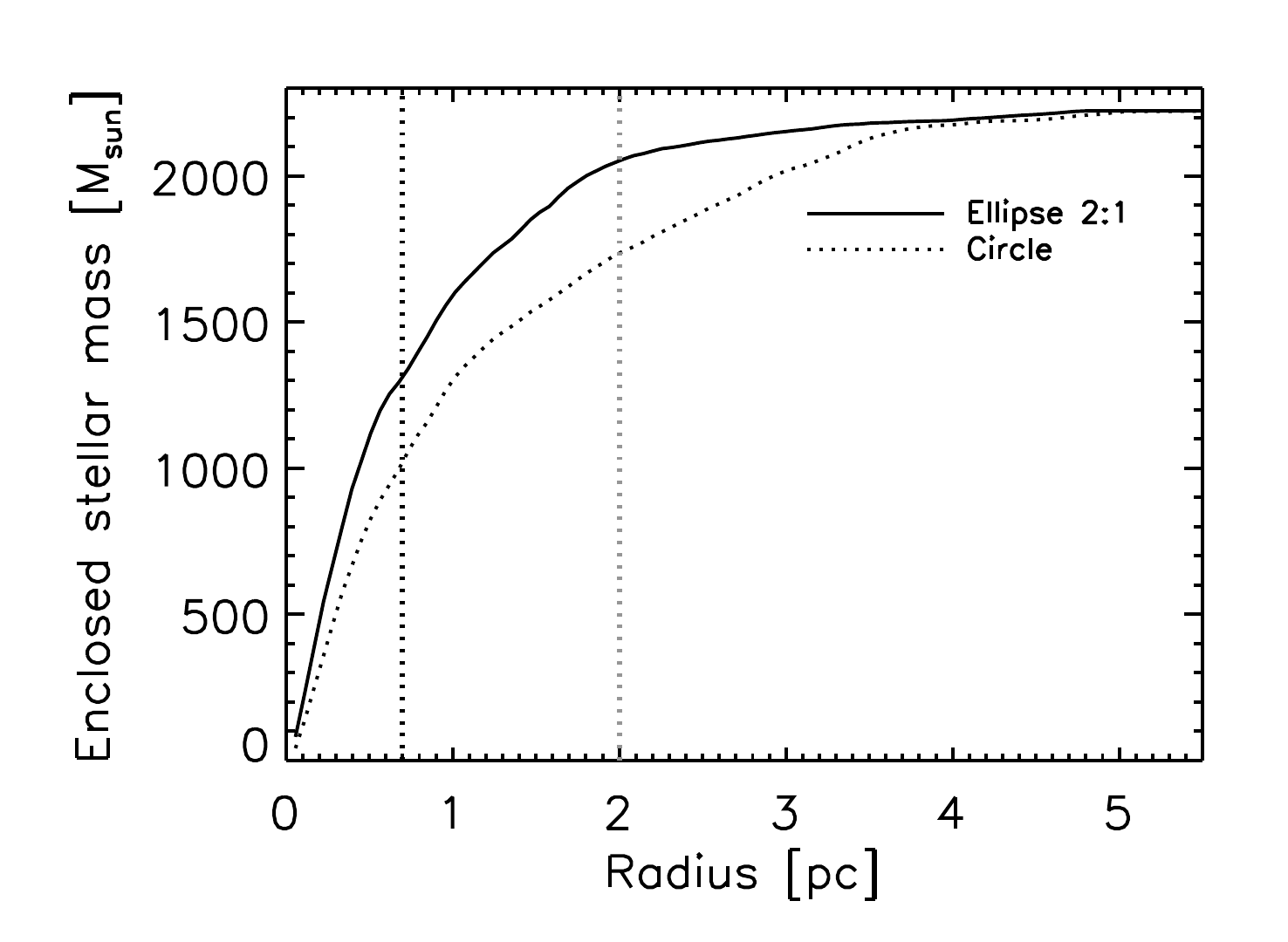}
  \caption{Stellar mass inside projected radii for
    elliptical (solid line, 2:1 axis ratio with long axis oriented
    North-South) and circular apertures (dotted line).  Here the
    ``radius'' of the elipse corresponds to the minor axis of the
    elliptical aperture.  Vertical dashed lines indicate the extent of
    the main structural components of the ONC: a clear stellar cluster
    core with r$\sim$0.7~pc and a more extended but lower density 
    "halo'' extending to about r$\sim$2.0~pc. The total stellar mass
    associated with the ISF is about 2200~\msun. The cluster mass
    within r$\sim$0.7, 2.0 ~pc is about 1300, 2000~\msun.}
  \label{fig:smasse}
\end{figure}

\section{Stellar mass distribution of the ISF}

We begin by analyzing the stellar mass distribution as a function of
$\delta$ and projected radial distance ($w$) from the gas ridgeline in
Figure~\ref{fig:smassd}.  The highly non-uniform stellar mass
distribution is evident, and peaks approximately in the center of the
ISF at the location of the ONC. Furthermore, the vast majority of the
stellar mass is located within a projected radial distance of about
2~pc from the dust ridgeline, consistent with the \citet{stutz16}
measurement of $\langle r_\perp\rangle = 1.48\,\pc$ for the disk
stars with APOGEE radial velocity measurements.

By analyzing the stellar mass profiles as a function of $w$, centered
on the peak of the stellar mass distribution located at
$\alpha=$~05:35:16.49, $\delta=$~$-$05:22:54.73 we obtain a
measurement of the projected cluster structure.
Figure~\ref{fig:smasse} shows that the stellar mass distribution can
be separated into three regions, separated at $w
\simeq\,0.7,2.0\,$\pc.  Based on the above figures we conclude that:
\begin{itemize}
\item[1.] The star cluster ``core'' at the center of the ISF has a
  projected radius $\lesssim$\,0.7~pc, corresponding to the rich
  center of the ONC where it is it is difficult to differentiate
  between a spherical symmetry and an elongated (elliptical)
  structure.
\item[2.] The star cluster ``core'' has minimal departures from
  circular symmetry below a projected radius $\lesssim$\,0.5~pc.
\item[3.] The ONC has a mass of $\sim\,$1000 to 1300~\msun\ within a radius of 0.7~pc.
\item[3.] The total stellar mass in the ISF is $\sim\,$2200~\msun.
\item[4.] There are $\sim\,$900~\msun\ of stars outside $r =$\,0.7~pc,
  distributed in a lower density stellar ``halo''.
\item[5.] These "halo stars" have a distribution that surrounds the
  main central cluster, but follows the gas filament (i.e., is
  highly elongated).
%%\item[6.] Outside the stellar ``halo'' practically no stars are found.
\end{itemize}

In particular, we emphasize that in the center of the cluster, at
radii $\lesssim$\,0.5~pc, Figure~\ref{fig:smasse} shows that the
departures from circular symmetry are minimal.  That is, the
differences between the elliptical and circular apertures in the
cumulative mass profiles are small, reaching 27\% at $r = 0.5$\,pc.
These differences are small compared to the overall variations in the
density profile (see below).  Thus the assumption of spherical
symmetry is reasonable.

\begin{figure}
  \scalebox{0.61}{\includegraphics{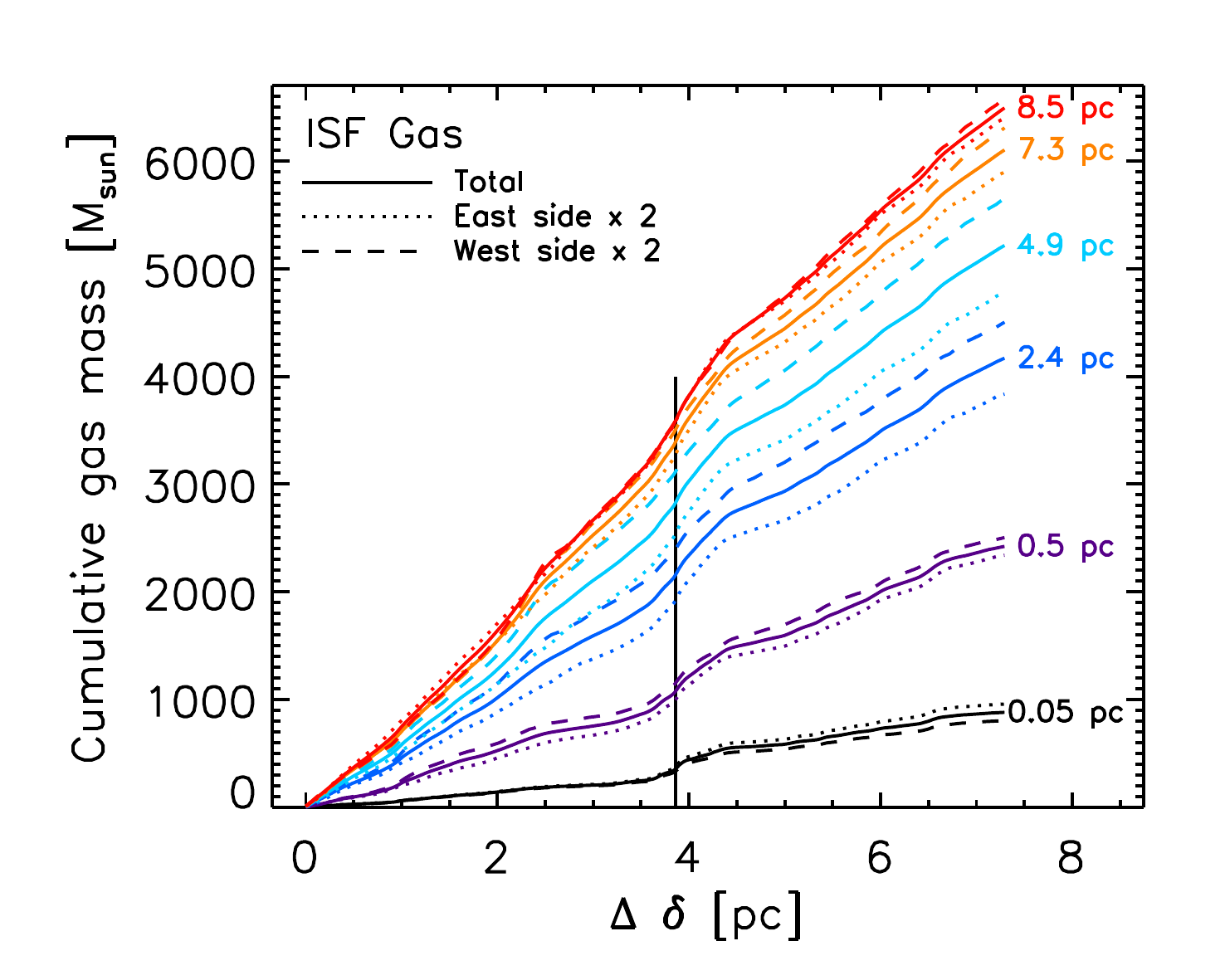}}
  \scalebox{0.61}{\includegraphics{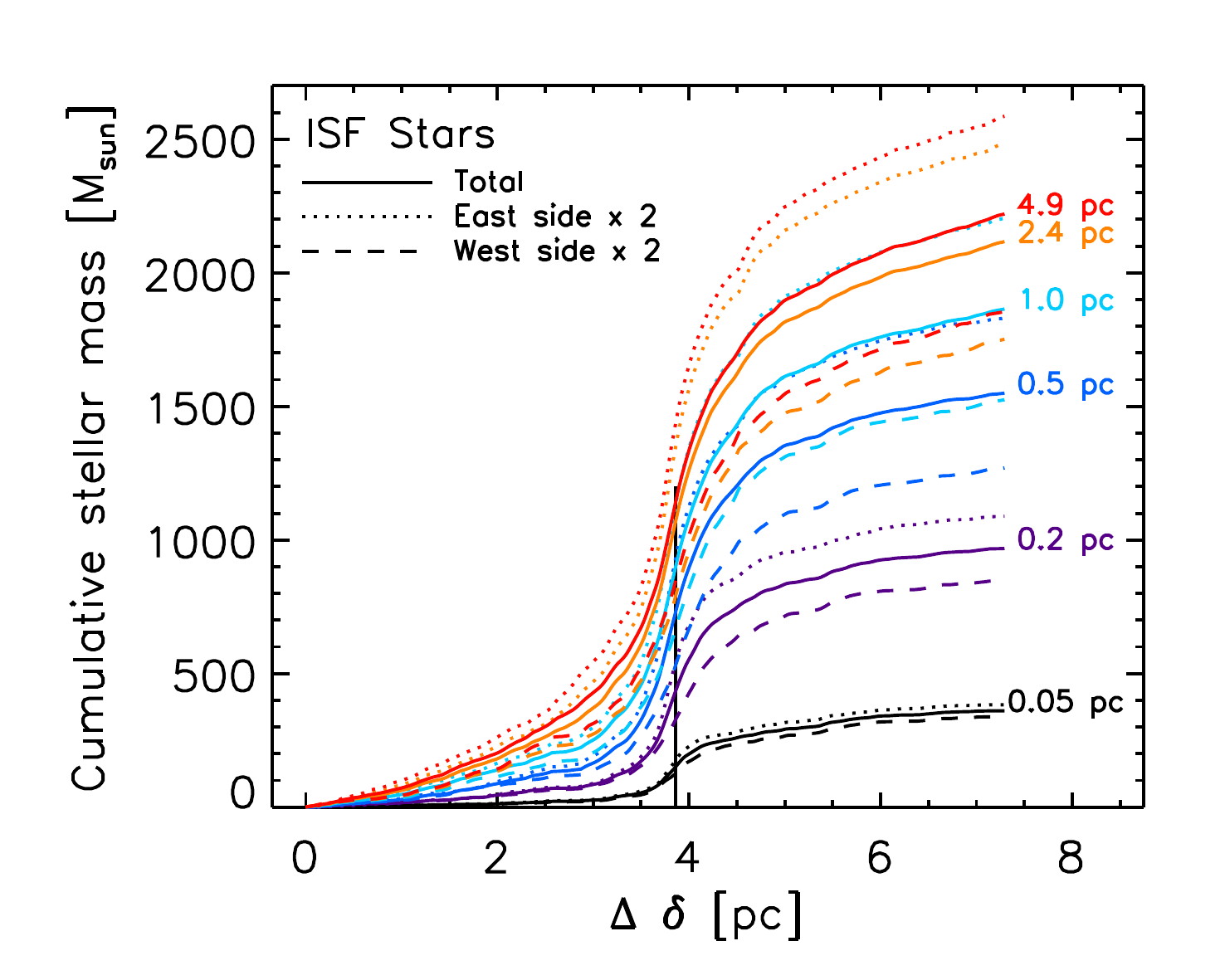}}
 \caption{Cumulative distributions of gas ({\it top}) and stellar
   ({\it bottom}) mass within various projected separations $w$ from
   the dust column density ridgeline for the ISF.  The cumulative
   distributions start at the southern boundary of the ISF (Dec =
   $-5.9\degree$).  The separate cumulative distributions for the east
   $(-w<x<0)$ and west $(0<x<+w)$ are shown in different line types.
   For ease of comparison (and clarity) these are multiplied by
   two. The vertical black line indicates the $\delta$ of the stellar
   center of the ONC.}
    \label{fig:gscum}
\end{figure}

\begin{figure}
  \scalebox{0.61}{\includegraphics{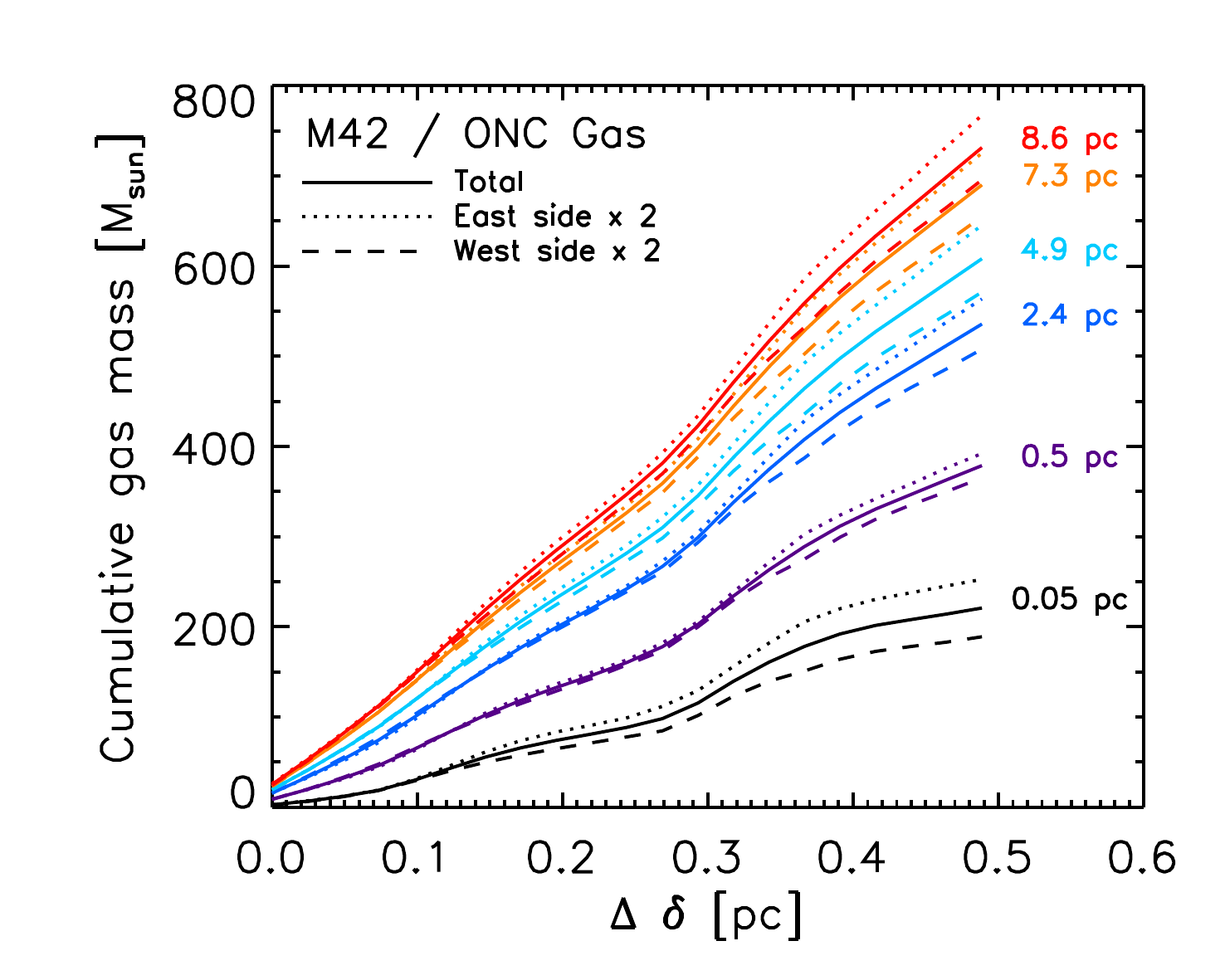}}
  \scalebox{0.61}{\includegraphics{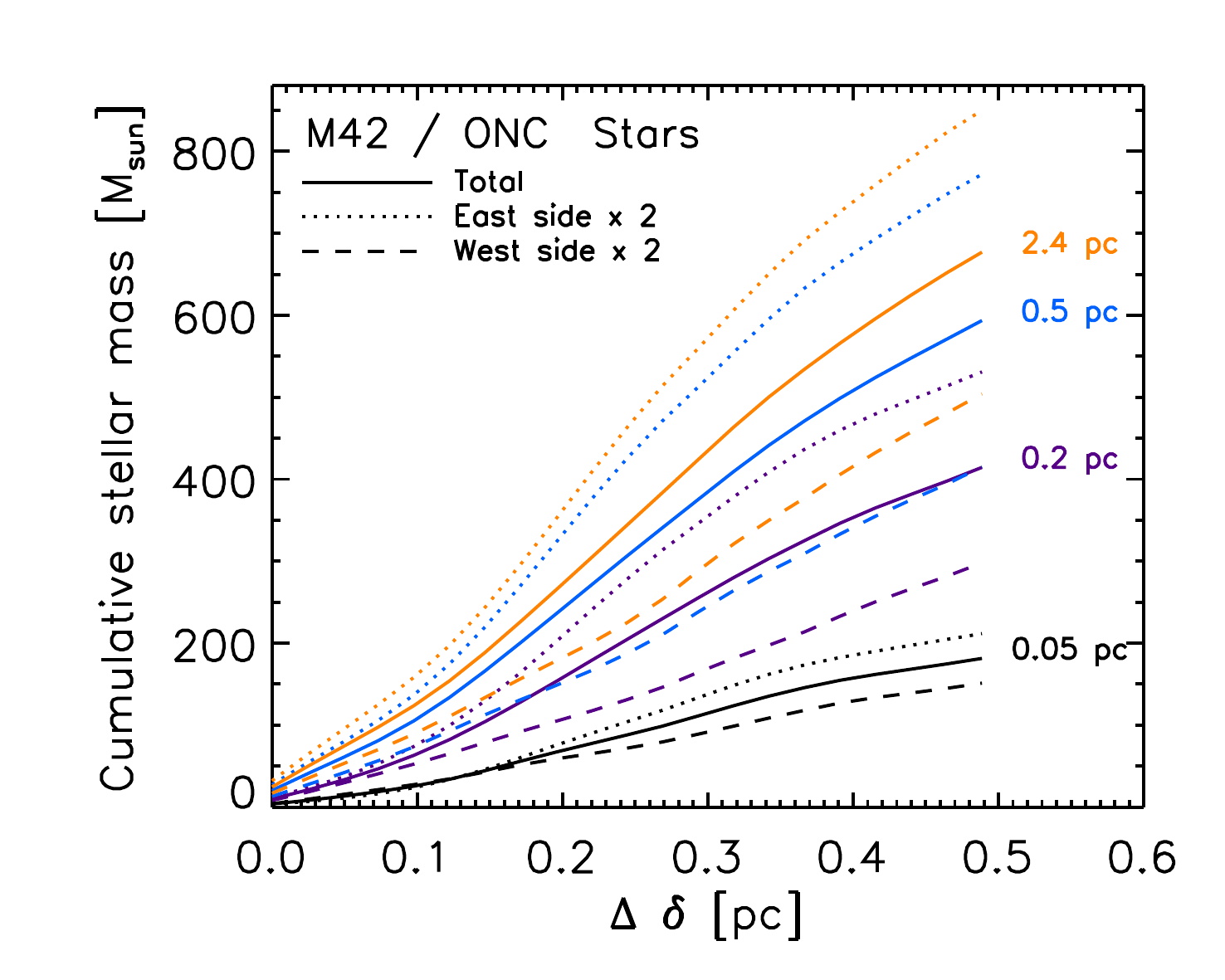}}
  \caption{Same as Figure~\ref{fig:gscum} zoomed in on the central 
    0.5~pc of the ONC.}
  \label{fig:gscumonc}
\end{figure}

\section{ONC gas and star volume density, gravitational potential, and gravitational field} 

It is clear from Figure~\ref{fig:gscum} that the gas (top panel) and
stellar mass (bottom panel) distributions are radically different.
The ISF stars exhibit a large mass concentration in the center of the
gas filament in the ONC \citep[e.g.,][]{megeath16}.  This stands in
contrast to the gas mass distribution, which is very close to uniform
along the ISF, as \citet{stutz16} showed.  In addition, this figure
shows that the correction to the saturated portion of the Herschel
N(H) image has a negligible effect on the gas mass distribution
profiles of the ISF (see Figure~4 of Stutz \& Gould 2016 for the ISF
mass distributions without the APEX 870~\micron\ correction).  We note
that previous works have characterized the projected stellar surface
density distributions of the ONC \citep[][; see also Gutermuth et al.\
2015, 2009 for methodology]{megeath16}\nocite{gutermuth09,gutermuth05}
and show that the stellar distribution has departures from circular
symmetry on scales $\gtrsim$\,0.5~pc.  Here we adopt the approximation
of spherical symmetry in the case of the stars in order to execute a
simple volume density analysis.  This assumption is valid on small
scales near the stellar cluster core, and not on larger scales as has
been previously shown (see above references).  Thus,
Figures~\ref{fig:smasse} and \ref{fig:gscum} motivate scrutinizing a
narrow range in $\delta$ within which the stellar mass distribution is
far more concentrated and a circular (or spherical) approximation is
reasonable.

\begin{figure}
  \includegraphics[width=\columnwidth]{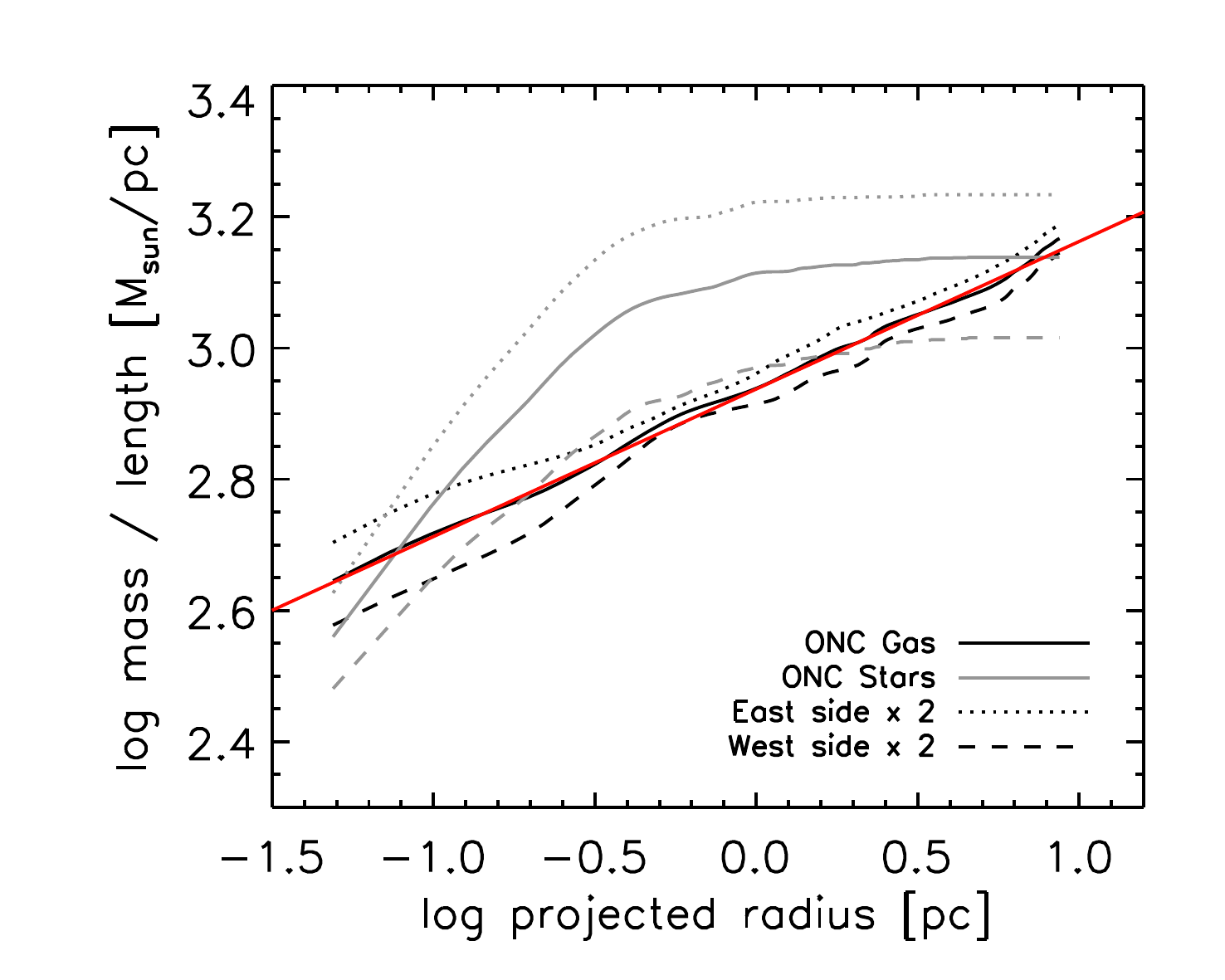}
  \caption{Cumulative mass/length as a function of enclosing width $w$
    integrated over a width in $\delta$\,=\,0.5~$\pc$ centered on the
    ONC. The ONC gas (black) and stars (grey) are shown separately.
    As in Figure~\ref{fig:gscumonc}, the east and west subsets are shown
    separately (after multiplying by 2). The red line indicates a
    power law of slope $\gamma$\,=\,0.225 (see
    eqn.~\ref{eqn:lambdag}).}
  \label{fig:cummasse}
\end{figure}

\begin{figure*}
  \scalebox{1.2}{\includegraphics{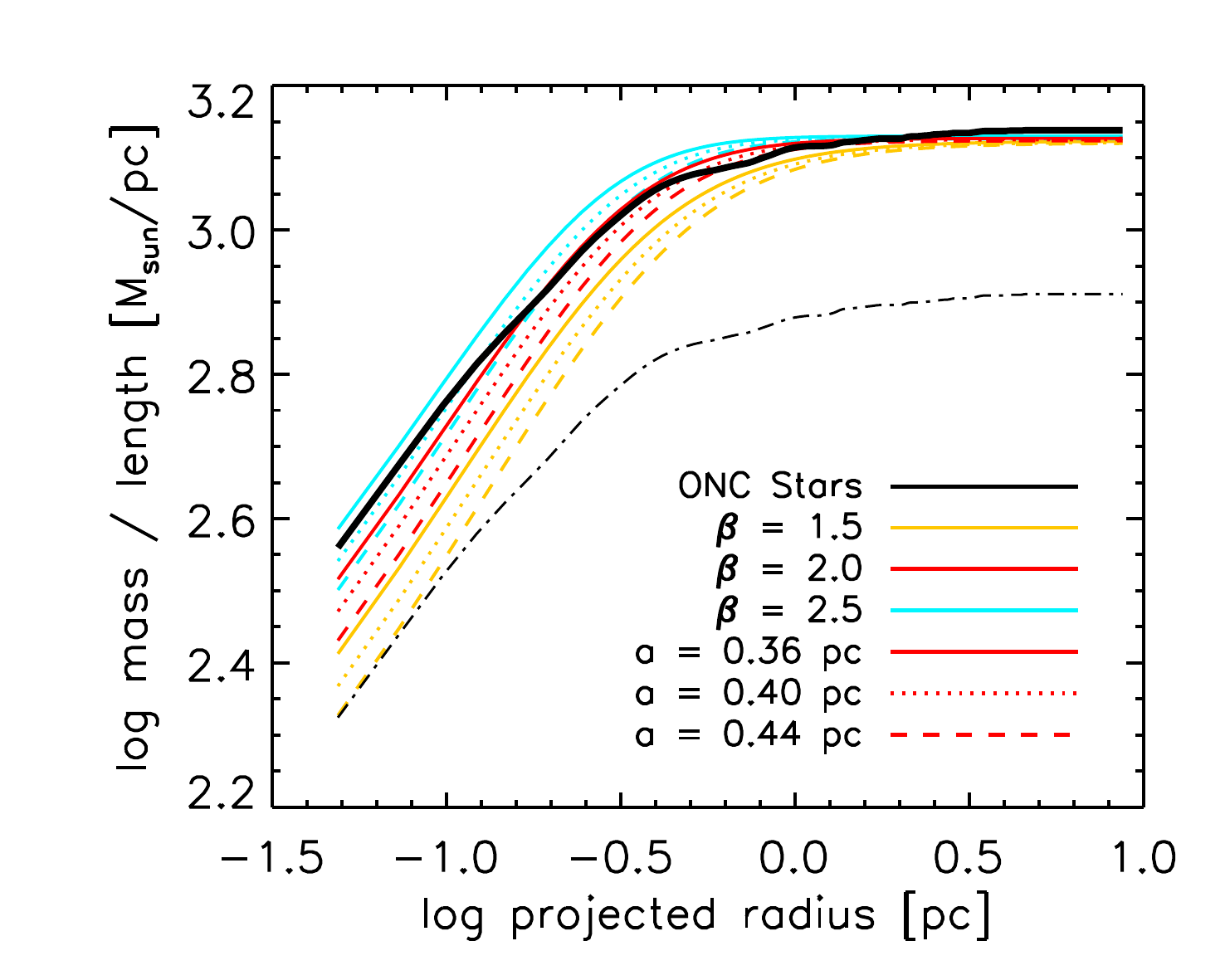}}
  \caption{Model fits to the stellar mass distribution in
    Fig.~\ref{fig:cummasse} using equation~\ref{eqn:surf}.  The
    best-fit model to the stellar distribution corresponds the
    red-dashed line, that is, to $a = 0.36$~pc and $\beta = 2$ (i.e.,
    a Plummer profile).  Here we adopt
    $M_{slice}/L = 1285\,\msun/{\rm pc}$ at $r = 1\,{\rm pc}$ as the
    total mass per unit length for the fit (see text). The dot-dashed
    line corresponds to the \spitzer star counts when ignoring the
    effects of nebulosity \citep[see text and ][]{megeath16}.}
  \label{fig:smassfit}
\end{figure*}

To this end we analyze the cumulative gas and stellar mass
distributions centered on the ONC star cluster, shown in
Figure~\ref{fig:gscumonc}. Here we analyze a much smaller range in
$\delta$ of 0.5~pc, indicated with boxes in Figures~\ref{fig:isf_cor}
and \ref{fig:isf_stars}.  Both the gas and star mas distributions are
calculated relative to the dust ridgeline \citep[see above
and][]{stutz16}. The top panel of Figure~\ref{fig:gscumonc} shows the
gas mass distribution over the center of the ONC, while the bottom
panel shows the stellar mass distribution.  From this figure we can
see that the cumulative gas mass increases relatively smoothly across
the 0.5~pc extent of the ONC region that we are scrutinizing, although
not as smoothly as for the ISF as whole \citep[see
Figure~\ref{fig:gscum} and][]{stutz16}. Meanwhile, the stellar mass
distribution exhibits significant curvature associated with underlying
circular (or spherical) structure of the cluster.

Taking advantage of the simple geometry implied by the two
distributions (cylindrical for gas and spherical for stars) we
construct mass per unit length profiles for both, shown in
Figure~\ref{fig:cummasse}.  Here, the projected mass per unit length
($M/L$) profiles are integrated over a height $\delta = 0.5$\,pc
centered on the region of highest projected stellar density, the
center of the ONC.  In practice this represents the $M/L$ profile over
a thin ``orange slice'' centered on the equatorial plane of the
stellar cluster.  This thin slice approach allows us to analyze the
gas and star distributions in the direction perpendicular to the gas
filament using simple geometric considerations.  In particular, this
approach allows us to ignore the more complex geometry of the star
distribution on larger scales, where elongation along the gas filament
precludes an accurate analysis using spherical or azimuthal averaging.

\subsection{ONC Gas}

Assuming cylindrical geometry, the gas distribution follows a power law of the form
\begin{equation}
\lambda(w) = K\biggl({w\over\pc}\biggr)^\gamma;
\qquad K = 866\,{M_\odot\over\pc},
\qquad \gamma = {0.225}.
\label{eqn:lambdag}
\end{equation}
Here $w$ is the projected radius as observed on the plane of the sky
(or the impact parameter to the dust ridgeline).  This power law has a
different normalization and index than that of the ISF filament as a
whole\footnote{\citet{stutz16} measure values of
  $K = 385\,M_\odot\,\pc^{-1}$ and $\gamma = 3/8$ (see their Eqn.~4)
  for the ISF as a whole.}  \citep{stutz16}.  We discuss this
difference below.  Such a power law line mass distribution can be
easily converted to a gas volume density profile assuming cylindrical
symmetry following Eqn.~5 of \citet{stutz16}:
\begin{equation}
\rho(R) = 25.9\,{M_\odot\over\pc^3}\biggl({R\over\pc}\biggr)^{\gamma-2}. 
\label{eqn:rho}
\end{equation}
Following \citet{stutz16} we also obtain the enclosed gas
line density $\Lambda(R)$, acceleration $a(R)$, and the gravitational
potential profiles:
\begin{equation}
\Lambda(R) = f(\gamma)\lambda(R) = 0.83\lambda(R) = 723\,{M_\odot\over\pc}\biggl({R\over\pc}\biggr)^\gamma,
\end{equation}
\begin{equation}
a(R) = 2G\Lambda/r = 6.2\,{\rm km\,s^{-1} Myr^{-1}}\biggl({R\over\pc}\biggr)^{\gamma-1}, 
\end{equation}
and
\begin{equation}
\Phi(R) = \eta(\gamma)G\lambda(R) = 27.6\,({\rm km\,s}^{-1})^2 \biggl({R\over\pc}\biggr)^{\gamma}.
\end{equation}
Here, $f(\gamma)$ and $\eta(\gamma)$ are defined in \citet{stutz16},
and have values of 0.834 and 7.418 respectively for $\gamma = 0.225$.

\subsection{ONC stars}

\begin{figure}
  \includegraphics[width=\columnwidth]{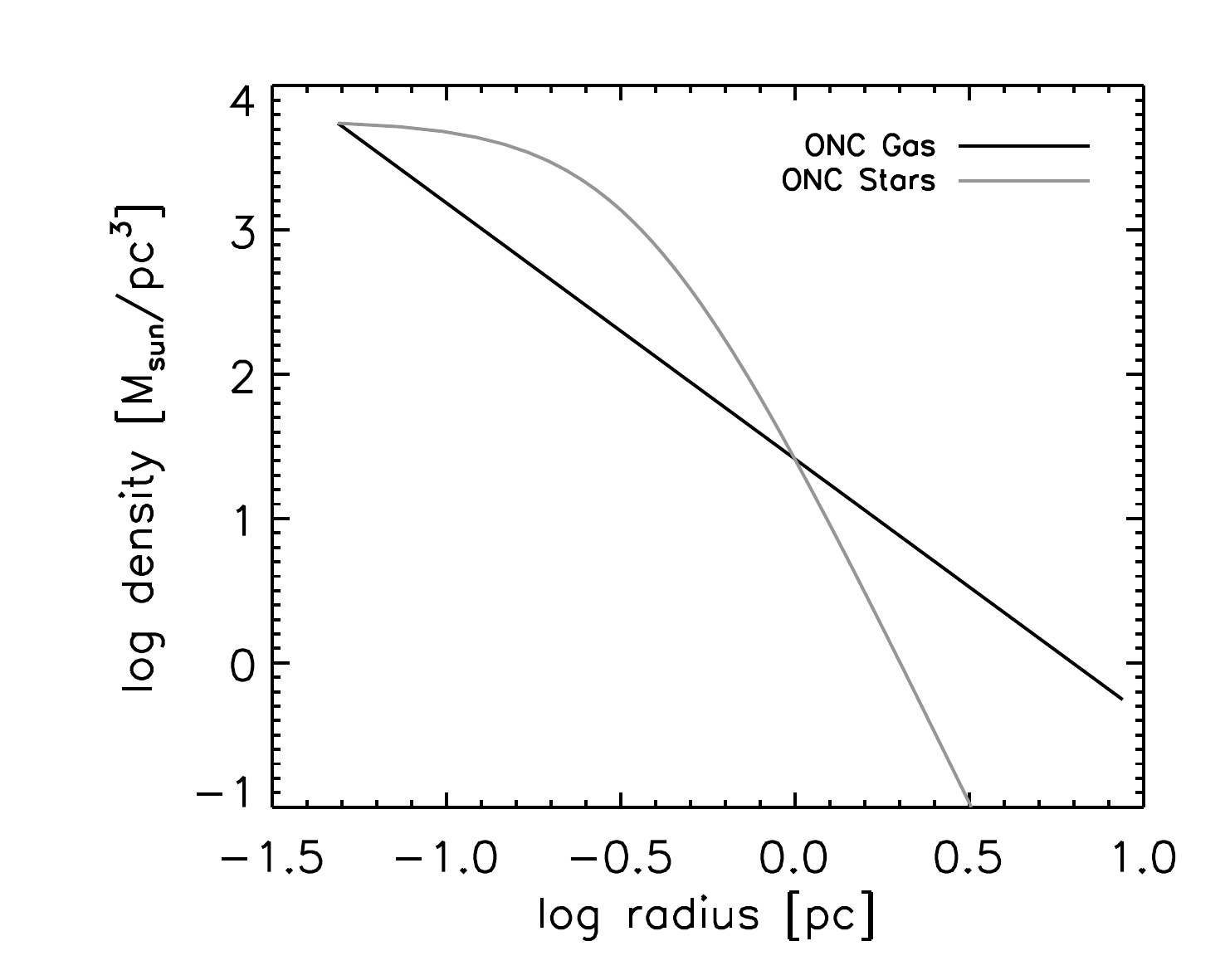}
  \caption{Volume density profiles of gas (black curve) and stars
    (grey) corresponding to the cumulative distributions shown in
    Figure~\ref{fig:cummasse}.}
  \label{fig:vol}
\end{figure}

Assuming spherical symmetry in the stellar core, we determine the stellar
volume density using a model of the form
\begin{equation}
\Sigma_s(b) \propto (1+(b/a)^2)^{-\beta} 
\label{eqn:surf}
\end{equation} 
to analyze the projected stellar $M/L$ profile
(Fig.~\ref{fig:cummasse}).  This is a generalized form of the
projected surface density of a Plummer profile; a value of $\beta = 2$
gives the familiar Plummer profile \citep{plummer11,binney08}, which
has a well defined density-potential pair, and is commonly used to
characterize star cluster profiles
\citep[e.g.][]{port2010,bianchini17}.  In Figure~\ref{fig:smassfit} we
show various $M/L$ curves for different values of both $\beta$ and
$a$.  The best match to the observed stellar distribution is given by
$\beta = 2$ and $a = 0.36$\,pc.  All models in
Figure~\ref{fig:smassfit} are normalized at large radii
($r = 1\,{\rm pc}$ with ${M_{\rm slice}/L} = 1285\,\msun/{\rm pc}$,
see below).  The normalization at small radii then strongly
discriminates against most models.  No model fits the data perfectly.
In particular, those that fit well at intermediate projected radii are
slightly too low at very small projected radii.  Two models apear to
fit the data at intermediate radii, $(\beta,a)=(2.0,0.36\,{\rm pc})$
and $(\beta,a)=(2.5,0.44\,{\rm pc})$.  Since
$(\beta,a)=(2.0,0.36\,{\rm pc})$ fits at the intermediate radii almost
perfectly and is closer to the data at small radii, we adopt this
model as a good analytic representation of the stellar profile.

From simple analytic analysis, we show in Appendix~A that for the case
of $\beta = 2$, we obtain a relation between $M/L$ measured in a thin
slice and the volume density profile, where
\begin{equation}
\rho_s(r)  =  K(1+(r/a)^2)^{-\gamma}; \qquad \gamma = \beta + 0.5,
\end{equation}
and
\begin{equation}
K = {3\over 2\pi a^2} {M_{\rm slice}\over L}\sqrt{1 + (L/2a)^2}.
\end{equation}
Here, ${M_{\rm slice}/L} = 1285\,\msun/{\rm pc}$ is the value of the
stellar $M/L$ measured from Figure~\ref{fig:smassfit} at $r = 1\,$pc.

The final volume density profile of the stars is 
\begin{equation}
 \rho_s(r)  =  5755\,{\msun \over {\rm pc}^3}(1+(r/a)^2)^{-5/2}.
\end{equation}
The corresponding gravitational acceleration and potential profiles are
\begin{equation}
a_s(r) = 37.3\,{\rm km\,s^{-1} Myr^{-1}} (1+(r/a)^2)^{-3/2} r/a,
\end{equation}
and
\begin{equation}
\Phi_s(r) = -13.4\,({\rm km\,s}^{-1})^2 (1+(r/a)^2)^{-1/2}.
\end{equation}
The mass and density profiles are consistent with previous
investigations into the density structure as a whole, for which e.g.,
\citet{dario14} find that a slope of $\gamma \sim 2.2$ for the volume
density profile.  In agreement with \citet{hillenbrand98}, we find an
inner flattening of the core density profile.  Below $r \sim
0.36\,{\rm pc}$ the stellar suface density flattens, and is
inconsistent with a power-law distribution that would continue into
the center of the cluster.  In terms of the density profile, the
values of $\beta$ and $a$ do not depend on the overall normalization
of the $M/L$ profile.  For comparison to \citet{megeath16}, in
Figure~\ref{fig:smassfit} we show the $M/L$ profile for the mass
profile based on the maps neglecting the nebulosity correction (see
Section~2).  The shape of the $M/L$ profile is the same, but the
normalization is a factor of 1.7 lower, with a value of ${M_{\rm
    slice}/L} = 746.7\,\msun/{\rm pc}$.  Thus, the parameters for the
shape of the volume density profile would not be affected.  However,
the nebulosity correction is essential for obtaining accurate stellar
profiles.  Therefore in what follows we use the fully corrected
profiles.

\begin{figure}
  \includegraphics[width=\columnwidth]{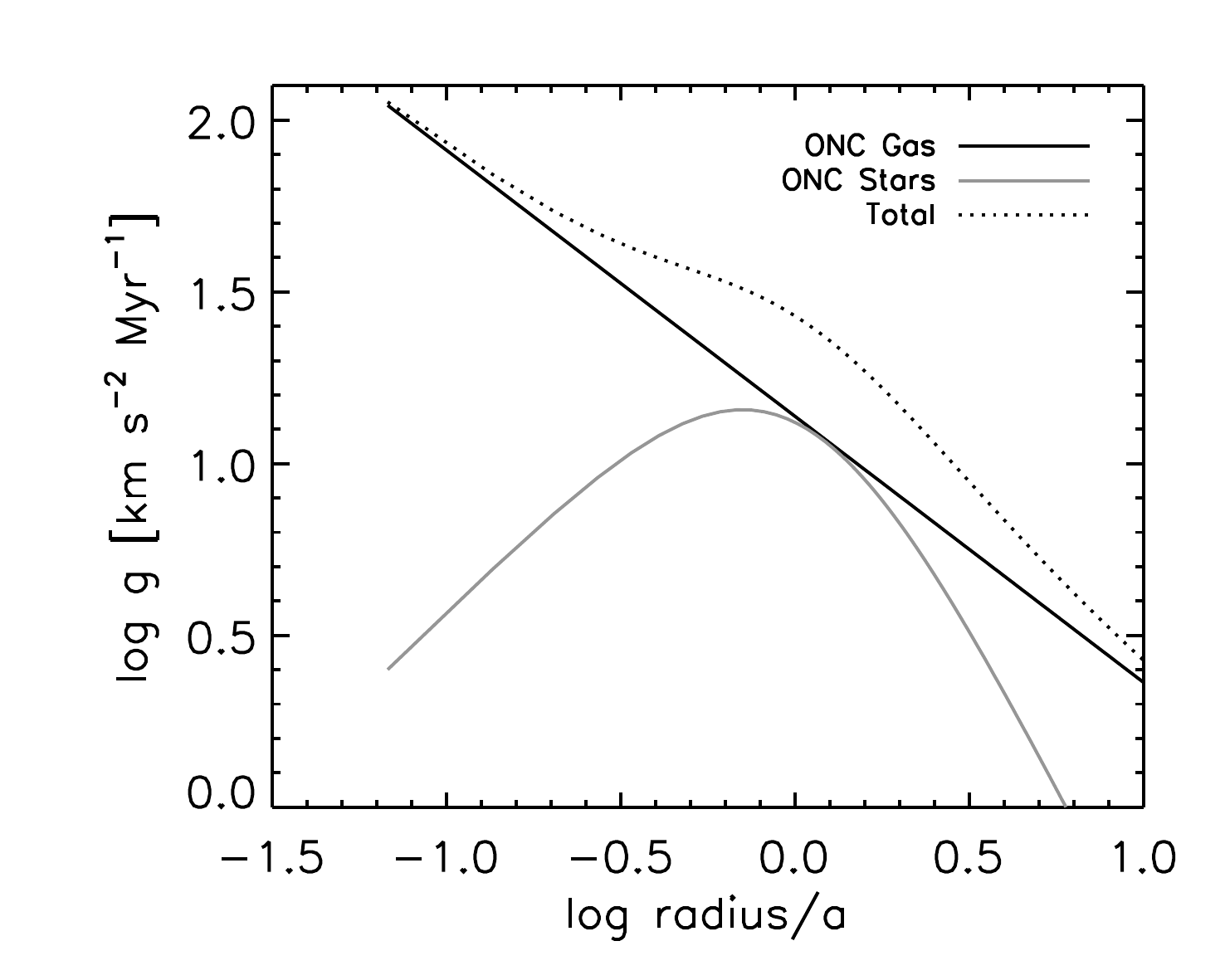}
  \caption{Gravitational field of gas (black curve) and stars (grey
    curve) corresponding to the density profiles shown in
    Figure~\ref{fig:vol}. The gas and star profiles reach
    near-equality at $r=a=0.36pc$, while the gravitational field due
    to the gas dominates everywhere else.}
  \label{fig:gf}
\end{figure}

\section{Discussion and conclusions}

In Figure~\ref{fig:vol} we show the volume density profiles for the
gas (black curve) and stars (grey curves).  One of the most striking
features of this diagram is that the stellar density surpasses the gas
inside r\,$\sim$\,1~pc, but outside the system is gas dominated.  In
Figure~\ref{fig:gf} we show the corresponding gravitational field
profiles for the gas, stars, and total. The stars contribute a maximum
to the field at $r/a\,\sim$\,0.8 ($\sim$\,0.3~pc), and have about
equal contribution as the gas near $r/a\,\sim$\,1.1 ($\sim$\,0.45~pc).
That is, the stars contribute most significantly to the field at
$r\,\sim\,a$, while the gas dominates at all other radii, both smaller
and larger.

We estimate the cluster crossing time
$t_{cross} = 2a / \sigma_{\rm obs} \sim 0.55$~Myr, where $a = 0.36$~pc
and $\sigma_{\rm obs} = 2.6$~km~s$^{-1}$.  We measure a velocity
dispersion for the stars of $\sigma_{\rm obs} = 2.6$~km~s$^{-1}$ from
the APOGEE data \citep[e.g.,][]{stutz16,dario16}.  Then we must
estimate the size of the cluster, i.e.\ the radius of a sphere
containing the stars having a velocity $\sim\,2.6$~km~s$^{-1}$. This
can be obtained from the data, for example from Figure~\ref{fig:vol}
where we can estimate a characteristic radius (containing most of the
stars) as $2a$.  If the lifetime of the ONC is $\sim$2-3~Myr
\citep{dario16}, then the core is about 4-6 crossing times young.  On
this relatively short timescale we do not expect significant internal
evolution to have taken place in the cluster. Furthermore, Stutz \&
Gould (2016) estimate the timescales of the gas filament motion of
$\sim\,0.6$~Myr, which is very similar to the cluster $t_{cross} $.

The ensemble of evidence presented here strongly suggests that the
cluster profile and dynamics are controlled by the gas filament.
First, the gravitational field is everywhere dominated by the gas.
This fact alone, established here for the first time, demonstrates
that the stellar cluster cannot be considered a virialized system
dominated by its own gravity.  Second, the transition from cluster
core to halo (at softening radius $a$) occurs at exactly the point
that the cluster gravity would begin to dominate that of the gas.
This indicates that the gas potential somehow limits the growth of the
cluster core.  That is, the structural parameters of the star cluster
might be determined by the gas filament.  Third, the internal cluster
timescale is the same as the oscillation timescale of the filament,
which is measured on spatial scales that are an order of magnitude
larger than the cluster.  

The coincidence of timescales, together with the fact that the
filament oscillations are clearly independent of the cluster, implies
that it is these oscillations that are driving the internal dynamics
of the cluster.  The driver for the gas oscillations is rooted in the
interaction between the magnetic field and gravity of the gas
filament.  That is, the magnetic field is not felt by the stars, they
move independently, only affected by gravity, both their own and that
of the gas.  Yet the gas does feel the magnetic field. The gas is
distributed in a $\sim$\,7~pc long filament which is much larger and
more massive than the star cluster, and which has a wave-like
morphology (and kinematics) that indicates that the gas is
oscillating, as proposed by Stutz \& Gould (2016).  In this picture,
the gas oscillations are determined by the interaction of the magnetic
field and the gravitational potential of the gas (which completely
dominates over the potential of the stars, except in a small region
near $r=a$ in the center of the ONC).  The gas then has a causal
relationship to the stellar cluster, whereby the gas oscillations
imprint on the stellar cluster structure through changes in the
gravitational field (felt by the cluster) which are caused by the
accelerating gas filament.  To second order, the density enhancement
of the stars in the center of the ONC might affect the oscillation,
but based on the star to gas mass ratios in the region, this effect is
not dominant.  The gas filament, both its gravitational-magnetic
oscillations and the gravitational tides that it generates, naturally
explain all three of the above elements.

The first key point is tides represent the differential acceleration
on a particle relative to the center of mass of a system that is
moving through an external potential.  If the cluster were sitting in
a non-accelerating external potential, there would be no tides: the 
particles (stars) would simply reorganize themselves according to the
external potential and their own velocity dispersion (somewhat
modified by their self-potential, which is subdominant).  On the other
hand, if the gas filament is accelerating due to non-gravitational
(i.e., magnetic) forces, then the cluster (which is only subject to
gravitational forces) will always be moving relative to the gas and so
will be subject to tides. In this way, the filament potential sets the
structural parameters of the ONC.  These tides will then automatically 
suppress low density structures, such as the outskirts of the cluster
\citep[for spherical analogs see ][]{lamers05,gieles16,renaud11}.

As shown by \citet{stutz16}, the stars are born (as protostars) on the
filament and have essentially zero velocity relative to the filament.
That is, the specific kinetic energy of the protostars is $\sim$\,6
times smaller than that of the stars.  On the other hand, the stellar
velocity dispersion is comparable to the velocity amplitude of
filament oscillations, which is one of the pieces of evidence that led
\citet{stutz16} to conclude that the stars were being ejected from the
filament as they mechanically decoupled from the gas.  This ejection
mechanism naturally explains why the cluster has a core: the core is
populated by stars that have been ejected (and fall into the
 potential well of the cluster) over a range of velocities
and positions as the filament oscillates.  Moreover, this naturally
explains why the crossing time of the stars is similar to that of the
filament oscillations.

Next, we should ask why there is a cluster at all if the kinematics
and potential are dominated by the filament?  The cluster forms first
of all because of a whip-like action of the oscillating filament,
which induces a shock, and consequent rapid concentration of gas, near
the filament's termination point at any given time.  NGC\,1981
\citep{pillitteri13} was formed at the end of the Orion A filament in
this way during one such whip-like snap about 4~Myr ago, and NGC\,1977
\citep{peterson08} was formed in the next whip-like snap about 2~Myr
ago.  Each of these cluster formation episodes truncated the Orion~A
filament progressively further to the south.  These are the only two
such episodes that are recognizable from the fossil record, but
undoubtedly there were earlier ones that occurred as Orion A gradually
retracted itself from Orion~B further to the north.  The ONC is the
result of the third, and currently ongoing, such snap that we can
recognize.  Further evidence that the ONC is the product of a
whip-like snap is provided by the N$_2$H$^+$ map of \citet[][; see
also Tatematsu et al.\ 2008]{hacar17}.  This shows that the
high-density gas is not simply concentrating to form stars, but is
also breaking into pieces just $\sim\,0.5$~pc to the south of the star
cluster.  These data also show that the dense gas is converging about
$\sim\,0.25$~pc to the south of the center of the cluster, and thus
far outside the gravitational center of the cluster, which may
be further evidence that the gas velocities reflect larger-scale
filament dynamics rather than infall into the star cluster.  

However, while the cluster does not dominate the gravitational field
in the direction perpendicular to the filament at any radius (see
Fig.~\ref{fig:gf}), it provides most of the restoring force in the
direction parallel to the filament.  We say ``most'' because the
filament is actually denser in the neighborhood of the cluster than in
other parts of the ISF, and so does provide some restoring force in
this direction.

Hence, in this picture, the gas continuously pours into the ``snapping
point'' of the filament, which is determined by its global structure,
including oscillations.  The conversion of gas into stars \citep[acting
as ``sink particles'', e.g., ][]{bate95} reduces the local pressure,
which drives a continuous replacement by new gas.  Because the stars
are born along the filament, and are confined by the gas potential
along the directions transverse to the filament, their self-gravity
can build up to the point that it restricts their dispersal along the
filament.

How does the gas filament cut off growth of the cluster core just at
the point that the cluster's gravity is beginning to dominate?  Again,
the key lies in the filament's oscillations.  As long as the nascent
proto-cluster remains a basically passive accumulation of stars, its
collective impact on the filament likewise remains minimal.  But to
the extent that the cluster becomes self-gravitating and capable of
following its own inertial orbit, it begins to impact the gas flows
within the filament as the filament oscillates away from it.  Indeed,
there is some evidence from the extinction measurements \citep[e.g.,
][]{odell01,wen95} that the cluster center is currently displaced from
(somewhat in front of) the gas filament.

This picture gives insight not just into the past history of the ONC,
but also its future.  Just as individual protostars separate from the
filament at the moment when their self-gravity enables them to
marginally decouple mechanically from the gas, so the ONC
proto-cluster as a whole will separate from the filament as it becomes
marginally self-gravitating and can decouple gravitationally from the
filament.  In the case of the individual protostars, their exit from
the filament initiates the next phase of their evolution as
disk-bearing stars that have been cut off from external gas accretion.
The proto-cluster's exit likewise initiates a new phase of life as a
nascent cluster that is cut off from the aggregation of new stars.  

In a richer environment than Orion, perhaps several such marginally
bound nascent clusters would merge to form a truly self-gravitating
system.  This will not be the fate of the ONC, however.  Rather,
because it will exit the filament when it is only marginally
self-gravitating, it will lose a large fraction of its stars and
become a shadow of its former self, similar to NGC\,1977 and
NGC\,1981.  In the still more distant future, less than 10 Myr, the
ONC will likely completely disperse, similar to the fate of the
clusters that formed before NGC\,1981 as the gap opened up between the
Orion~A and Orion~B filaments.

Future numerical simulations coupling oscillating filaments to N-body
dynamics \citep[][Matus Carrillo et al., in prep.]{boekholt17} of star
clusters will stringently test this hypothesis.  Furthermore, ALMA
observations of the number and distribution of protostars within the
ONC will provide essential observational constraints on the origin of
the stars populating the cluster.

\section{Acknowledgments}
We thank the referee for their helpful comments which aided in the
clarity of the presentation. We thank Andrew P.\ Gould, Tjarda C.\ N.\
Boekholt, Dominik R.\ G.\ Schleicher, Michael Fellhauer, Xiaoling
Pang, and S.\ Thomas Megeath for many stimulating discussions.  We are
very grateful to S.\ Thomas Megeath for providing the \spitzer star
count data in map form.  We are grateful for the generous hospitality
of Shanghai Normal University (SNU) where a portion of this work was
carried out.  We acknowledge funding from the ''Concurso Proyectos
Internacionales de Investigaci\'on, Convocatoria 2015'' (project code
PII20150171) and the BASAL Centro de Astrof\'isica y Tecnolog\'ias
Afines (CATA) PFB-06/2007.  This paper includes data from \herschel, a
European Space Agency (ESA) space observatory with science instruments
provided by European--led consortia and with important participation
from NASA.  We include data from the Atacama Pathfinder Experiment
(APEX), a collaboration between the Max-Planck-Institut f\"{u}r
Radioastronomie, the European Southern Observatory, and the Onsala
Space Observatory.

%%\bibliographystyle{mnras}
%%\bibliography{ref}

\appendix

\section{From an $M/L$ to a volume density}

We assume polytrope density and projected density profiles of 
the form 
$$
\rho(r) \equiv K(1+(r/a)^2)^{-\gamma};
\qquad
\Sigma(b) \equiv C(1+(b/a)^2)^{-\beta}.
$$
Here, 
$$
\Sigma(b) = \int_{-\infty}^\infty dz\,\rho(\sqrt{z^2+b^2})
= 2\int_0^\infty dz\,K((a^2 + b^2)/a^2 + z^2/a^2)^{-\gamma}
$$
$$
= 2(1+(b/a)^2)^{-\gamma}\int_0^\infty dz\,K(1 + z^2/(a^2+b^2))^{-\gamma}
$$
$$
= 2(1+(b/a)^2)^{-\gamma}\sqrt{a^2+b^2}\int_0^\infty {dz\over\sqrt{a^2+b^2}}
\,K(1 + z^2/(a^2+b^2))^{-\gamma}
$$
$$
= 2K(1+(b/a)^2)^{-\gamma+0.5}a\int_0^\infty dx
\,(1 + x^2)^{-\gamma}.
$$
So, $\beta=\gamma-0.5$ and $C=2Ka\int_0^\infty dx\,(1 + x^2)^{-\gamma}$
where
$$
\int_0^\infty dx\,(1 + x^2)^{-\gamma}
=\int_0^{\pi/2}d\theta\cos^{2\gamma-2}\theta
$$
$$
=0.5(-1/2)!(\gamma-3/2)!/(\gamma-1)!
$$
which for $\gamma=5/2$ is $0.5(-1/2)!/(3/2)! = 2/3$.
That is, $C = (4/3)aK$.

Now consider an ``orange slice'', of infinitesimal width, but
offset from the center by $q$:
$$
\lambda(q) = \int_{-\infty}^\infty dy\Sigma(\sqrt{q^2+y^2})
= 2\int_0^\infty dy\, C((q^2+a^2+y^2)/a^2)^{-\beta}
$$
Then, using exactly the same algebraic steps as above
$$
\lambda(q) = \kappa(1 + (q/a)^2)^{-\delta}
$$
where $\delta=\beta-0.5$ and $\kappa=Ca(-1/2)!(\beta-3/2)!/(\beta-1)!$

Finally, the total mass in a thick slice within $\pm L/2$ of the center is
$$
M_{\rm slice}=\int_{-L/2}^{L/2} dq\,\kappa(1+(q/a)^2)^{-\delta}
=2\kappa a\int_0^{L/2} {dq\over a}\,\kappa(1+(q/a)^2)^{-\delta}
$$
$$
= 2\kappa a\int_0^{L/2a} dx(1+x^2)^{-\delta}
= 2\kappa a\int_0^{\tan^{-1}(L/2a)} d\theta\cos^{2\delta -2}\theta
$$
$$
= 2Ka^3{[(-1/2)!]^2(\gamma-3/2)!(\beta-3/2)!\over(\gamma-1)!(\beta-1)!}
\int_0^{\tan^{-1}(L/2a)} d\theta\cos^{2\delta -2}\theta
$$
$$
= 2Ka^3{[(-1/2)!]^2(\beta-1)!(\beta-3/2)!\over(\beta-1/2)!(\beta-1)!}
\int_0^{\tan^{-1}(L/2a)} d\theta\cos^{2\delta -2}\theta
$$
$$
= {2\pi Ka^3\over\delta}
\int_0^{\tan^{-1}(L/2a)} d\theta\cos^{2\delta -2}\theta
$$
This can be evaluated numerically, but for the special case of $2\delta=$''integer'',
it can be done analytically.   In particular, for $\gamma=5/2$, i.e.,
$\delta=3/2$, we obtain 
$$
\int_0^{\tan^{-1}(L/2a)} d\theta\cos^{2\delta -2}\theta
=\sin(\tan^{-1}(L/2a))={1\over \sqrt{1 + (2a/L)^2}}. 
$$

That is,
$$
M_{\rm slice} = {4\pi K a^3\over 3}\,{1\over \sqrt{1 + (2a/L)^2}}.
$$
Note that for the special case of $L=\infty$, we recover the standard
formula for the cluster mass.  For the special case of a thin
slice (which was previously considered) we get
$M_{\rm slice} = 2\pi K a^2 L/3$, i.e., proportional to $L$.
Motivated by this ``$M/L$'' form, we can write the
general case as
\begin{equation}
M_{\rm slice} = {2\pi K a^2 L\over 3}\,{1\over \sqrt{1 + (L/2a)^2}}.
\label{eqn:a1}
\end{equation}
From this,we can see that for the particular case of
$(L,a)=(0.5,0.36)\,{\rm pc}$, the ``adjustment factor'' relative to the
previous ``thin slice'' reasoning is $1/\sqrt{1+(0.5/0.72)^2}=0.82$.

The net result is that for our parameters,
$\beta=2.0$, $a=0.36\,$pc, $L=0.5\,$pc,
$$
K = {3\over 2\pi a^2} {M_{\rm slice}\over L}\sqrt{1 + (L/2a)^2}
= {4.48\over {\rm pc}^2}{M_{\rm slice}\over L}.
$$

\begin{figure}
  \includegraphics[width=\columnwidth]{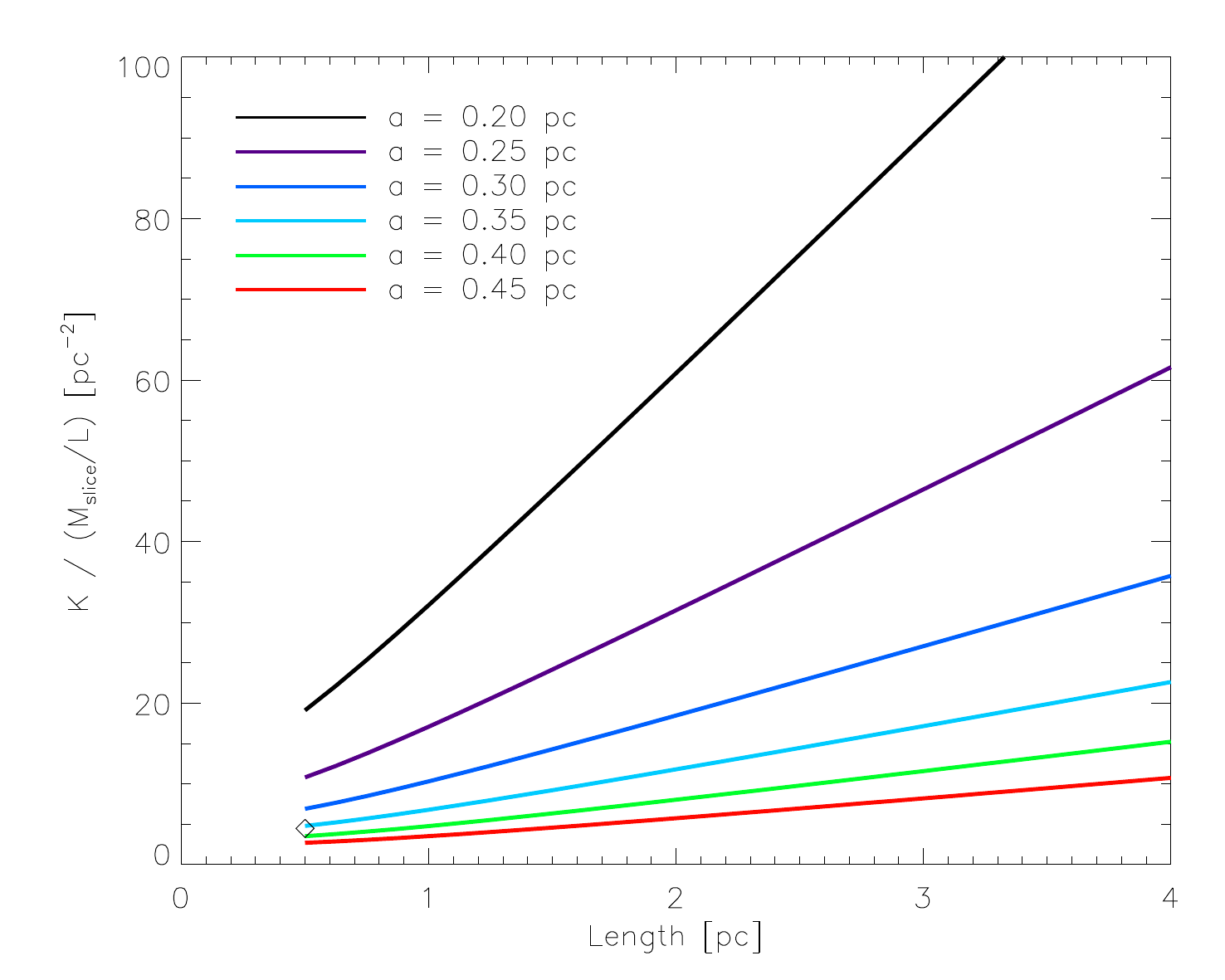}
  \caption{Relationship between slice thickness (L), Plummer profile
    softening scale (a), total mass per unit length ($M_{\rm slice}/L$),
    and the normalization of the Plummer density profile shown in
    equation~\ref{eqn:a1}.  The diamond symbol indicates the
    particular case for the ONC.}
  \label{fig:kml}
\end{figure}

\bsp
\label{lastpage}
\end{document}